\documentclass[superscriptaddress,prd,twocolumn,nofootinbib,notitlepage,showpacs]{revtex4-1}

\usepackage{bm,bbm,caption,subcaption,graphicx}
\usepackage{amsmath,amsfonts,amssymb,times,psfrag}
\usepackage{bbold}
\usepackage{mathrsfs}

\DeclareMathAlphabet{\mathpzc}{OT1}{pzc}{m}{it}
\newcommand{\beq}{\begin{eqnarray}}
\newcommand{\eeq}{\end{eqnarray}}

\begin{document}

\title{Quantum critical diffusion and thermodynamics in Lifshitz holography}
\author{Brandon W. Langley}
\email{blangle2@illinois.edu}
\author{Udit M. Gupta}
\email{uditmg2@illinois.edu}
\author{Philip W. Phillips}
\email{dimer@illinois.edu}
\affiliation{Institute of Condensed Matter Theory and Department of Physics\\University of Illinois at Urbana-Champaign, Urbana, IL 61801}

\begin{abstract} 
We present the full charge and energy diffusion constants for the Einstein-Maxwell dilaton (EMD) action for Lifshitz spacetime characterized by a dynamical critical exponent $z$.  Therein we compute the fully renormalized static thermodynamic potential explicitly, which confirms the forms of all thermodynamic quantities including the Bekenstein-Hawking entropy and Smarr-like relationship. Our exact computation demonstrates a modification to the Lifshitz Ward identity for the EMD theory. For transport, we target our analysis at finite chemical potential and include axion fields to generate momentum dissipation.  While our exact results corroborate anticipated bounds, we are able to demonstrate that the diffusivities are governed by the engineering dimension of the diffusion coefficient, $[D]=2-z$.  Consequently, a $\beta$-function defined as the derivative of the trace of the diffusion matrix with respect to the effective lattice spacing changes sign precisely at  $z=2$.   At $z=2$, the diffusion equation exhibits perfect scale invariance and the corresponding diffusion constant is the pure number $1/d_s$ for both the charge and energy sectors, where $d_s$ is the number of spatial dimensions. Further, we find that as $z\to\infty$, the charge diffusion constant vanishes, indicating charge localization. Deviation from universal decoupled transport obtains when either the chemical potential or momentum dissipation are large relative to temperature, an echo of strong thermoelectric interactions.
\end{abstract}

\maketitle

\section{Introduction}

Because most condensed matter systems do not conform to the full Lorentz symmetry and contain dynamical behavior characterized by Lifshitz transitions \cite{Bianconi2018, Norman2010, Shi2018, Volovik2018}, tailoring the AdS/CFT program to condensed matter systems such as the cuprates requires a considerable extension.  The simplest proffer to engineer such a non-relativistic setup is a Lifshitz geometry characterized by a dynamical critical exponent $z$ \cite{Kachru2008,Horava2009,Balasubramanian2011, Kiritsis2013, Jensen2018}. A metric ansatz that encodes the features of both the dynamical exponent and a finite temperature is the form
 \begin{align}
	ds^2 = \frac{dr^2}{r^2 f(r)} - r^{2z} f(r) dt^2 + r^2 d\vec{x}^2_{d_s},
\end{align}
which defines a horizon by the largest root of $f(r_+) = 0$ and boundary at $r \to \infty$ where $f \to 1$. This ansatz encapsulates the scaling symmetry
\begin{align}
	r \to c^{-1} r, \quad t \to c^z t, \quad x^i \to c x^i.
\end{align}
at the boundary. For $z \neq 1$, this metric ansatz cannot be a vacuum solution to the Einstein equations. Indeed, such a Lifshitz geometry requires a nontrivial bulk stress-energy tensor.  Herein lies the problem: there is no unique way of engineering the requisite stress-energy tensor.  
 
A full analytic solution to an asymptotically Lifshitz geometry that features a black hole can be constructed via an Einstein-Maxwell-dilaton (EMD) action. This model is a direct extension of the AdS-Reissner-Nordstr\"{o}m black hole to $z \neq 1$. This action is well-known in the literature as it has served as the workhorse for most of the Lifshitz papers \cite{Gouteraux2011, Fan2013, Gouteraux2014, Tarrio2012, Kiritsis2015, Kiritsis2017, Papadimitriou2014}.

Our work addressed the open problem of the construction of the transport properties in a fully-renormalized solution to the EMD action that features both a chemical potential and a set of spatially-dependent axion fields that induce momentum dissipation for general $z$ and dimensionality. The action is addressed completely at the level of the static background and DC transport. This range of analysis allows us to compute the set of both static susceptibilities and conductivities under a uniform ``Lifshitz charge", which by the Einstein relations obtain the full set of thermoelectric diffusion constants. We are able to compare and contrast our results with universal features of diffusivity by Blake \textit{et al.} \cite{Blake2016, Blake2017}. They consider the limits of decoupled charge and thermal diffusion, and we find exact agreement in this limit. However, we find deviations from such behavior when matter interactions are cranked such that the thermoelectric coupling is significant.

A significant result we obtain once all the dust settles is that the length dependence of the transport properties, although they are governed by several independent scales ranging from the chemical potential to the strength of momentum dissipation, are ultimately controlled by the engineering dimension $[D]$ of the diffusion constants;  by inspection of the diffusion equation, $[D]=2-z$.  Consequently, the effective $\beta$-function \cite{Anderson1979},
\begin{align}
	\beta\equiv\frac{\partial}{\partial \ell}\frac{(\textrm{tr} \bm{D} )T}{v_B^2},
	\label{betaf}
\end{align}
should exhibit universal features as a function of the characteristic length scale $\ell$. In Eq. \eqref{betaf}, we measure the diffusion matrix $\bm{D}$ against the characteristic scale $v_B^2/T$, defined in terms of the butterfly velocity and the temperature \cite{Blake2016}. We find that $\textrm{sgn}(\beta) = \textrm{sgn}(2-z)$, indicating a fixed point at $z=2$.  At the scale-invariant point $z=2$,
 diffusion is given exactly by the dimensionless number
\begin{align}
	D(z=2)=\frac{1}{d_s}.
\end{align}
Our diffusion constants are strictly positive unlike the previous results with restricted range of validity for $z$ \cite{Pang2010,Ge2018}. We find $z=2$ corresponds to the fixed point associated with the length dependence of the diffusivities, in direct analogy with the $\beta$-function in Anderson localization \cite{Anderson1979}.  Our conclusion here is made possible entirely because we have a regularizable boundary theory.  

We find in general that diffusivity bounds \cite{Hartnoll2014,Hartman2017} do indeed exist for Lifshitz holography, with two possible violations. The first for $z =1$, as is standard, has a divergent energy diffusion constant in the absence of momentum dissipation, caused by the inability of any sourced momentum to relax. The second occurs at $z \to \infty$, whereupon charge becomes completely localized and does not diffuse, manifest in the vanishing of the upper bound in the respective diffusion constant.
\begin{figure}[t!]
	\centering
	\includegraphics[scale=0.4]{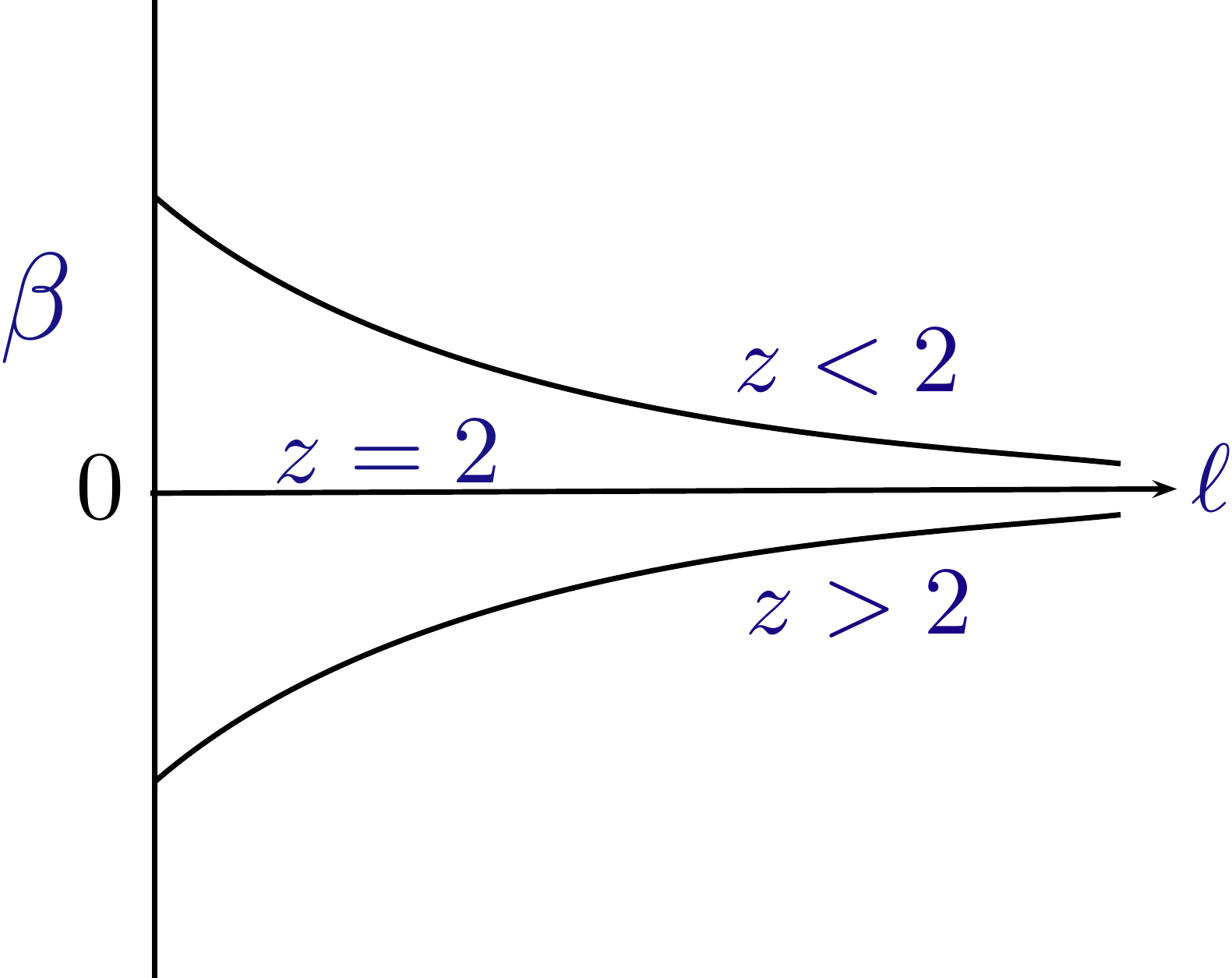}
	\captionsetup{justification=raggedright,singlelinecheck=false}
	\caption{Heuristic plot of the $\beta-$function defined as the derivative of the trace of the diffusion matrix with respect to the system size.  At $z=2$, the diffusion coefficient is a universal dimensionless constant determined solely by the number of spatial dimensions.  Away from $z=2$, the diffusion constants (either charge or energy) have opposite slopes relative to an increase in the system size indicating a fixed point at $z=2$.  Regardless of $z$, $\beta$ asymptotes to zero as the system size increases, indicating the bounded nature of the diffusivities. At $z=2$, the diffusion equation is scale-invariant.}
	\label{BI}
\end{figure}

As a final important note, our exact treatment finds that the Ward identity associated with the Lifshitz symmetry in the EMD model. We find that the boundary stress-energy tensor $\mathbbm{T}^a_{\;\;b}$ and dilaton response $\lambda_\phi\mathbbm{O}_\phi$ obey
\begin{align}
	z \mathbbm{T}^t_{\;\;t} + \mathbbm{T}^{x^i}_{\;\;x^i} + \lambda_\phi \mathbbm{O}_\phi = 0,
\end{align}
which aligns with the encoding of the Lifshitz symmetry via dilatation. This is a slight contrast to the usual identity which omits any contribution from the dilaton. In spite of this, these results are not contradictory in the context of \cite{Kiritsis2015, Kiritsis2017} for instance, due to an alternative construction of the boundary theory and the dual stress-energy tensor. In fact, due to the nebulous nature of the interpretation of the boundary geometry, there is some leeway in the formulation.

\section{An Abridged History of Lifshitz Holography}

The short history of Lifshitz holography is one of many fits and starts. The initial advance \cite{Kachru2008} in this context supplemented the standard bulk Lagrangian with two gauge fields, a 1-form and a 2-form both, coupled together via a topological term that controls the dynamical exponent $z$.  Though a clean analytic construction, it is restricted to $d_s=2$ spatial dimensions and is not amenable to emblackening factors which would encode a horizon.

Several other models have been proposed and analyzed \cite{Cassani2011}, such as the Einstein-Proca model \cite{Andrade2013, Papadimitriou2016}. However the EMD model imbibes the most robust features for thermodynamics. One of the long-standing issues with this theory is the absence of a renormalization scheme for the boundary action, unlike AdS \cite{Balasubramanian1999}. Without such a scheme, there is no real interpretation to the response functions and thermodynamics of the system. While some models could be worked out under specific conditions ---  the $z=2$ Schr\"{o}dinger symmetry \cite{Chemissany2012} a case in point --- or certain response functions obtained such as the specific heat \cite{Keranen2012}, the general theory remained elusive \cite{Taylor2015}. One of the culprits is the $U(1)$ field responsible for turning on $z \neq 1$ is poorly behaved at the boundary. There have been several proposals for dealing with this divergence. The original proposition for handling this divergence was to perform a Legendre transformation \cite{Tarrio2012} to instead consider the stable dual $U(1)$ current as the fundamental variational instead of the electric field.

The Legendre transformation alone leaves the scheme incomplete. An alternative was proposed. Kiritsis and Matsuo's hydrodynamic ansatz, wherein the constant parameters controlling the static solution are promoted to slowly-varying functions of time and space, permits an analytic solution to the induced fluctuation equations. As its title suggests, this formulation allows all thermodynamic quantities to be expressed as components of a nonrelativistic fluid. The controlled expansion allows them to make contact with a renormalization scheme \textit{without} performing a Legendre transformation. Their solution involves an infinite series of counter-terms involving the divergent $U(1)$ field and supports slowly varying transport properties.

The next stride was made by Cremonini \textit{et al.} who examined the transverse modes of the EMD theory \cite{Cremonini2017, Cremonini2018}. Previous attempts on this model ignored the crucial coupling between the two $U(1)$ fields \cite{Alishahiha2012, Kuang2017}, which must be present for a nontrivial solution. Cremonini \textit{et al.} sought heat and charge transport response functions of the system at low frequencies. Therein, their program enabled a renormalization of specifically the transverse modes:  they (1) perform the Legendre transformation for the divergent $U(1)$ field, (2) build a $second$ ADM breakdown to separate time and space in the boundary with a timelike shift, and (3) renormalize the theory in terms of the corresponding $U(1)$ current and timelike shift. A feature of this scheme is that the counter-terms actually depend on both the non-normalizable and normalizable modes of the model. A renormalization scheme depending upon the theory's renormalizable modes is usually problematic, but in fact the counter-terms can be state-dependent for systems perturbed by irrelevant operators \cite{vanRees2011}.

\section{Action and Static Background}

We suppose the Einstein-Maxwell-dilaton (EMD) action
\begin{widetext}
\begin{align}\label{action}
	I = - \int_{\mathcal{M}}\sqrt{-g}\left( R - \frac{1}{2}(\nabla\phi)^2- V(\phi) -\frac{1}{4}\sum_{q=1}^2 Z_q (\phi) F_q^2 -\frac{1}{2}X(\phi)\sum_{I=1}^{d_s} (\nabla \chi_I)^2\right) -\int_\mathcal{\partial \mathcal{M}} \sqrt{-\gamma} 2K + I_{\text{c.t.}},
\end{align}
\end{widetext}
where $I_{\text{c.t.}}$ is a smattering of counter-terms to give us well-defined boundary action.  Here, $\mathcal{M}$ is a $d_s+2$-dimensional Lorentzian manifold and $\partial\mathcal{M}$ is its boundary. This action features two $U(1)$ fields, one which will serve to assist turning on a nontrivial $z \neq 1$ solution and one which generates a standard chemical potential $\mu$, a straight extension of the usual AdS-Reissner-Nordstr\"{o}m black hole.  The axion fields will generate momentum dissipation. This action yields the equations of motion
\begin{widetext}
\begin{subequations}
\begin{align}
	& E_{ab}=R_{ab}-\frac{1}{2}\nabla_a \phi \nabla_b \phi-\frac{1}{d_s}V(\phi)g_{ab} \nonumber \\
	& \qquad \quad -\frac{1}{2}\sum_{q=1}^2 Z_q(\phi)\left(F_{q,ac}F_{q,b}^{\;\;\;c}-\frac{1}{2d_s}F_q^2 g_{ab}\right)-\frac{1}{2}X(\phi)\sum_{i=1}^{d_s}\nabla_a \chi_i \nabla_b \chi_I = 0, \\
	& D_\phi = \Box\phi-V'(\phi)-\frac{1}{4}\sum_{q=1}^2 Z_q'(\phi)F_q^2-\frac{1}{2}X'(\phi)\sum_{I=1}^{d_s} (\nabla \chi_I)^2=0, \\
	& M_q^a = \nabla_b\left(Z_q(\phi)F_q^{ab}\right) = 0, \\
	& \Xi_I = \nabla_a\left(X(\phi)\nabla^a \chi_I\right) = 0.
\end{align}
\end{subequations}
\end{widetext}
This system has a static solution, with matter fields given by
\begin{widetext}
\begin{align}\label{staticsol}
	& \phi = \lambda_\phi \left( \ln r + \phi_1 \right), \quad \lambda_\phi = \sqrt{2d_s(z-1)}, \nonumber \\
	& V(\phi) = -(z+d_s-1)(z+d_s), \quad Z_q(\phi) = e^{2 \frac{\lambda_q}{\lambda_\phi} \phi}, \quad X(\phi) = e^{2 \frac{\lambda_\chi}{\lambda_\phi} \phi},					\nonumber\\
	& A_1 = \sqrt{\frac{2(z-1)}{z+d_s}} e^{d_s\phi_1}\left(r^{z+d_s}-r_+^{z+d_s}\right)dt, \quad \lambda_1 = -d_s, \nonumber\\
	& A_2 = \mu_2\left[1-\left(\frac{r_+}{r}\right)^{z+d_s-2} \right]dt, \quad \lambda_2 = z-1, \nonumber\\
	& \chi_I = k \delta_{Ii}x^i, \quad \lambda_\chi = -(z-1),
\end{align}
\end{widetext}
and the emblackening factor
\begin{align}
	& f(r) = 1 + \frac{z+d_s-2}{2d_s}\frac{\mu_2^2 e^{2(z-1)\phi_1}}{r_+^2}\left(\frac{r_+}{r}\right)^{2(z+d_s-1)} \nonumber \\
	& \qquad\quad +\frac{1}{2(z - d_s)}\frac{k^2 e^{-2(z-1) \phi_1}}{r_+^{2z}}\left(\frac{r_+}{r}\right)^{2z} - M\left(\frac{r_+}{r}\right)^{z+d_s},
\end{align}
where $M$ is the mass of the black hole,
\begin{align}
	& M = 1+\frac{z+d_s-2}{2d_s}\frac{\mu_2^2 e^{2(z-1)\phi_1}}{r_+^2}+\frac{1}{2(z - d_s)}\frac{k^2 e^{-2(z-1) \phi_1}}{r_+^{2z}}.
\end{align}
Note that this solution demands $z \geq 1$. For the critical case $d_s=z$, a logarithmic singularity obtains, and the emblackening factor takes the form
\begin{widetext}
\begin{align}
	& f(r) = 1+ \frac{z-1}{z} \frac{\mu_2^2 e^{2(z-1)\phi_1}}{r_+^2}\left(\frac{r_+}{r}\right)^{4z-2} \nonumber \\
	&\qquad\quad-\left[1 + \frac{z-1}{z} \frac{\mu_2^2 e^{2(z-1)\phi_1}}{r_+^2}-\frac{1}{2}\frac{k^2 e^{-2(z-1) \phi_1}}{r_+^{2z}}\ln\left( \frac{r_+}{r}\right)\right]\left(\frac{r_+}{r}\right)^{2z}.
\end{align}
\end{widetext}
For the analysis in this paper, we will avoid this singularity and forbid $k\neq 0$ for this critical case, as it inherently would not modify the thermodynamics after renormalization. The temperature is
\begin{widetext}
\begin{align} \label{temp}
	T = \frac{r_+^{z+1}f'(r_+)}{4\pi} = \frac{r_+^z}{4\pi}\left( z+d_s - \frac{(z+d_s-2)^2}{2d_s} \frac{\mu_2^2 e^{2(z-1)\phi_1}}{r_+^2} - \frac{1}{2} \frac{k^2 e^{-2(z-1)\phi_1}}{r_+^{2z}}\right),
\end{align}
\end{widetext}
derived by eliminating the conical singularity at the horizon.

\section{Renormalization}

In order to have a well-defined boundary action, and therefore dictionary, we must have a renormalization scheme to ensure all divergences are removed. If our action's variation can be expressed as
\begin{align}
	\delta I = \int_{\partial \mathcal{M}} \sum_n \Pi_n^{(\phi)}\delta \phi_n,
\end{align}
where we have symplectic data and the products of all the variations $\delta \phi_n$ and their radial  conjugate momenta $\Pi_n^{(\phi)}$ are $O(1)$ as $r \to \infty$, then we have ascertained a renormalized boundary action. This scheme can be implemented not requiring a full, generalized solution but rather utilizing details of specific solutions only as necessary. The scheme for the low-frequency transverse transport properties was first laid out in \cite{Cremonini2018}. If the reader seeks a more generalized approach beyond the scope of specific solutions, it is wise to turn to the radial Hamilton-Jacobi equations to define a boundary potential via a functional derivative expansion, as in \cite{Papadimitriou2011,Chemissany2015}. For our purposes, we need only examine the possible forms of nontrivial counter-terms to ascertain the renormalized action within the scope of our solution.

The variation of the action \eqref{action} yields
\begin{widetext}
\begin{gather}
	\label{boundvary}\delta I = \int_{\partial \mathcal{M}} \left( \frac{1}{2} T^{ab} \delta \gamma_{ab} - \sum_{q=1}^2 J_q^a \delta A_{q,a} + \mathcal{O}_\phi \delta \phi \right), \\
	T^{ab} = 2\sqrt{-\gamma}\left( K^{ab} - K \gamma^{ab}\right), \quad J_q^a = \sqrt{-\gamma}N_b  Z_q(\phi) F_q^{ab}, \quad \mathcal{O}_\phi = \sqrt{-\gamma} N_a \nabla^a \phi	,
\end{gather}
\end{widetext}
where $N^a$ is a unit vector normal to the boundary hypersurface foliating the bulk spacetime along the radial direction, and $K_{ab} \equiv \nabla_{(a} N_{b)}$. We have neglected the conjugate momenta for the axions as they vanish at the level of the static background and will not contribute to the DC currents. We absolutely do not have renormalized symplectic data, with the most glaring issue being that $A_1$ is divergent. However, its conjugate momentum is $O(1)$; hence we can switch to a stable scheme via a Legendre transformation. We select the counter-terms
\begin{widetext}
\begin{align}\label{counterterms}
	I_{\text{c.t.}} = \int_{\partial \mathcal{M}} \left( A_{1,a} J_1^a + c_{0}^{(\text{vol})}\sqrt{-\gamma} + c_{0}^{(J_1^2)} \frac{J_1^2}{\sqrt{-\gamma}Z_1(\phi)} + \cdots \right),
\end{align}
\end{widetext}
where we can see the first term switches our potential to vary under $J_1$ instead of $A_1$, changing the boundary condition from Dirichlet to Neumann, as proposed in \cite{Tarrio2012}.  We will thusly refer to this action as $I_N$. While $A_2$ and $J_2$ already combine to yield an $O(1)$ boundary contribution, we must yet ensure the other responses are renormalized.   Note that  $J_1^2$ counter-term is necessary for a finite response but it can also take the interpretation of a double-trace deformation.  Equipped with these counter-terms, we define the new responses as
\begin{widetext}
\begin{subequations}
\begin{align}
	& \mathbbm{T}^a_{\;\;b} = 2\sqrt{-\gamma}\left(K^a_{\;\;b}-K\delta^a_{\;\;b}+\frac{1}{2}c_{0}^{(\text{vol})}\delta^a_{\;b}\right)+\frac{2 c_{0}^{(J_1^2)}}{\sqrt{-\gamma}Z_1(\phi)}\left(J_1^a J_{1,b} - \frac{1}{2}J_1^2 \delta^a_{\;\;b} \right)+\cdots, \\
	& \mathbbm{A}_{1,a} = A_{1,a} + \frac{2 c_{0}^{(J_1^2)} J_{1,a}}{\sqrt{-\gamma}Z_1(\phi)}+\cdots, \\
	& \mathbbm{O}_\phi = \sqrt{-\gamma}N_a \nabla^a \phi- c_{0}^{(J_1^2)}\frac{Z_1'(\phi)}{Z_1(\phi)^2}\frac{J_1^2}{\sqrt{-\gamma}}+\cdots.
\end{align}
\end{subequations}
\end{widetext}
We find that
\begin{align}
	c_{0}^{(\text{vol})} = z + 2d_s -1, \quad c_{0}^{(J_1^2)} = \frac{1}{2(z+d_s)}
\end{align}
is sufficient. These counter-terms are in agreement with those used to analyze DC conductivities in \cite{Cremonini2018}. Note that implicit in the ellipses \textit{are} remaining counter-terms, but the outlined contributions are the only terms that contribute to the finite action. All other terms will simply cancel divergences without contributing to the free energy and are worked out accordingly in Appendix \ref{backren}.

If we substitute in our static solution, we will find as $r \to \infty$,
\begin{subequations}
\begin{align}
	& \mathbbm{T}^t_{\;t} = d_s M r_+^{z+d_s} + \cdots, \\
	& \mathbbm{T}^{x^i}_{\;\;x^j} = - M r_+^{z+d_s}\delta^i_{\;j} + \cdots, \\
	& \mathbbm{A}_{1,t} = \sqrt{\frac{2(z-1)}{z+d_s}}e^{d_s\phi_1}\left(\frac{M}{2} - 1 \right) r_+^{z+d_s} + \cdots, \\
	& \mathbbm{O}_\phi = - \sqrt{2d_s(z-1)} \frac{M}{2} r_+^{z+d_s} +\cdots.
\end{align}
\end{subequations}
 Also note that the stress-energy tensor is actually traceless, which is a bit surprising for a Lifshitz theory as this is usually indicative of a scale-invariant theory, though it is not a necessary condition. Regardless, the Lifshitz scaling symmetry is manifestly encoded as a dilatation. This is reflected in the new Ward identity,
\begin{align}
z \mathbbm{T}^t_{\;\;t} + \mathbbm{T}^{x^i}_{\;\;x^i} + \lambda_\phi \mathbbm{O}_\phi = 0.
\end{align}
This modification to the Ward identity for the anisotropic Weyl transformation \cite{Arav2017} is perhaps not so strange. Unlike the pure AdS case, there is no simple interpretation of the boundary metric. Under separate constructions such as the Newton-Cartan background, modifications are expected \cite{Hartong2015}.

While this renormalization machinery is sufficient for the static background, when it comes to the currents we will find that $z \neq 1$ causes the scaling of the source terms to be divergent. We can remove the source terms from our boundary action with a simple reformulation of our intrinsic metric,
\begin{align}
\gamma_{ab}dx^a dx^b = -n^2\left( dt - \frac{n_i}{n^2} dx^i\right)^2 + \sigma_{ij} dx^i dx^j,
\end{align}
which functions akin to an inverted ADM formalism, which we further discuss in Appendix \ref{adxRenorm}.

\subsection{Lifshitz Constraint}
	We note here that the renormalized $\mathbbm{A}_{1,t}$ has no independent free parameters: it is entirely determined in terms of the sources and vevs of the other operators in the theory.   As a result, we cannot source $A_{1,t}$ independently of the dilaton source $\phi_1$ or the energy density $\mathbbm{T}^t_{\;t}$. This is not an accident and is a consequence of a "second-class" constraint required to support Lifshitz asymptotics in EMD theories with massless U(1) gauge fields. This was shown using a radial Hamiltonian formulation in \cite{Chemissany2015} where they found the most general solution of the radial H-J equations. It was not a result of the equations of motion but instead is imposed by hand for asymptotically Lifshitz backgrounds, and therefore is a second-class constraint on the radial Hamiltonian system. 
	
	The space of asymptotic solutions with Lifshitz boundary conditions was identified along with all local counterterms required to regularize the divergent on-shell action as well as to identify the normalizable/non-normalizable modes in this theory. The second-class constraint leads to a fixed form for the asymptotic behavior of the gauge field $A_{1,t}(x)$, Eq. (5.26) in \cite{Chemissany2015} given by
	\begin{align}
		A_{1,t}(x) \approx \sqrt{\frac{2(z-1)}{(z+d_s)}} \,n(x)e^{d_s \phi_{1}(x)}(1 + r^{-\Delta_-}\psi_-(x)).
	\end{align}
	The asymptotic form is entirely determined in terms of the the modes $n(x)$, which source the energy density, and $\phi_{(1)}(x)$ which sources the dual dilaton operator. While there is no independent source for the dual operator to this gauge field it still has an independent, non-zero VEV which we can roughly interpret from the appearance of the free parameter $\psi_-(x)$ as a subleading term. Therefore, a consistent definition of the holographic dictionary only exists when we impose Neumann BCs on $A_{1,t}$ but not in the Dirichlet case as we do not have independent control over the chemical potential. This constraint also shows up in section VII of this paper when we consider perturbations of the conserved currents at the horizon. It forces us to set the current $j^{(\rho_1)} = 0$ in order to have a holographic interpretation of the horizon dynamics consistent with the UV theory. In so doing, $\nabla \mu_1$ is determined in terms of the other sources $\nabla T, \nabla\mu_2$ and hence cannot be independently varied. 
	\subsection{Double Trace Deformation?}
	If it were possible to independently vary the source term in $A_{1,t}$ (the chemical potential), we could in fact switch  to Dirichlet boundary conditions and consider the diffusion problem with $j^{(\rho_1)} \neq 0$. An attempt was made to vary the chemical potential $\mu_1$ by hand through the introduction of a double trace deformation,
	\begin{subequations}
	 \begin{align}
	 	I &\rightarrow I_{\beta} = I_{(0)} + \beta_0\int d^{d}x  f(\phi) \frac{(J_1^t)^2}{\sqrt{-\gamma}}\\
	 	A_{1,t} &= \frac{\delta I_\beta}{\delta J_1^t} = A^{(0)}_{1,t} - 2\beta_0 f(\phi) J_1^t\\
	 	\mu_1 &= \lim_{r\rightarrow \infty} A_{1,t} = \mu^{(0)}_1 - 2\beta_0 f(\phi_1) J_1^t.
	 \end{align}
	 \end{subequations}
	  Such an approach is misguided as this term is already included in $I_{c.t.}$ as a state-dependent counter term where its 'deformation parameter' is a fixed function of the state parameters $z,\phi_1$. 
	  \begin{align}
	  	\int d^dx\, \frac{c_{0}^{(J_1^2)}}{Z_1(\phi)}\frac{(J_1^t)^2}{\sqrt{-\gamma}} = \int d^dx\, \frac{e^{2d_s\phi_1}}{2(z+d_s)}\frac{(J_1^t)^2}{\sqrt{-\gamma}} 
	  \end{align}
	  One could interpret the emergence of such a term as an irrelevant deformation that induces an RG flow to a Lifshitz UV fixed point. Its appearance is also reflected in the holographic renormalization of Lifshitz EMD theories in \cite{Chemissany2015} where it shows up as a term in the general solution of the (homogeneous) H-J radial equations, which are also the most general set of local counterterms. If one were to perform an additional double trace deformation and follow its RG flow (roughly equivalent to solving the deformed radial HJ equations) it would spoil the Lifshitz asymptotics by flowing away from the Lifshitz UV fixed point.   Hence, such a double-trace deformation was not pursued here.

\section{Free Energy}
The free energy can be computed from the on-shell action. The Ricci scalar

\begin{align}
\begin{split}
	R &= \frac{1}{2}(\nabla\phi)^2+\left(1+\frac{2}{d_s}\right)V(\phi)+\\
	&+\left(\frac{1}{4}-\frac{1}{2d_s}\right)\sum_{q=1}^2 Z_q (\phi) F_q^2+\frac{1}{2}X(\phi)\sum_{I=1}^d (\nabla\chi_I)^2
\end{split}
\end{align}

can be plugged in to yield the on-shell bulk action
\begin{align}
	I_{\text{bulk}}^{(\text{o.s.})} = -\int_\mathcal{M} \sqrt{-g}\left(\frac{2}{d_s}V(\phi)-\frac{1}{2d_s}\sum_{q=1}^2 Z_q(\phi) F_q^2 \right),
\end{align}
which we can expressly integrate. Combining with our boundary terms, we find that the full on-shell boundary action is
\begin{widetext}
\begin{align}
	I_N^{(\text{o.s.})} =  \frac{W_N}{T} = \text{vol}_{d_s} \frac{ r_+^{z+d_s}}{T}\left( -z+\frac{(z-2)(z+d_s-2)}{2d_s}\frac{\mu_2^2 e^{2(z-1)\phi_1}}{r_+^2}+\frac{z}{2(z - d_s)}\frac{k^2 e^{-2(z-1)\phi_1}}{r_+^{2z}}  \right),
\end{align}
\end{widetext}
where $\text{vol}_{d_s}$ is a $d_s$-dimensional spatial volume and $W_N$ is the ``Neumann" free energy. We are working in an ensemble with $W_N(T, \phi_1, \mu_2,k)$, indicating our independent variables, with $r_+(T, \phi_1, \mu_2, k)$ are an implicit function that solves Eq. \eqref{temp}.  A priori, we notice that the charge density $J_1^t = \sqrt{2(z-1)(z+d_s)}e^{-d\phi_1}$ is a function of only $\phi_1$, meaning that the parameter $\phi_1$ directly and single-handedly sources both the responses $\mu_1$ and $\mathbbm{O}_\phi$.

Let us compute (and verify) our thermodynamic quantities. First, the entropy,
\begin{align}
	\mathcal{S} & = -\frac{\partial W_N}{\partial T}, \nonumber\\
	& = \text{vol}_{d_s} 4\pi r_+^{d_s},
\end{align}
where $\text{vol}_{d_s} r_+^{d_s}$ is exactly the surface area of the black hole. This is the celebrated Bekenstein-Hawking relation (where we have chosen $G_N = \frac{1}{16\pi}$).
Second, the charge associated with the chemical potential $\mu_2$ is
\begin{align}
 	\mathcal{Q}_2  & = - \frac{\partial W_N}{\partial \mu_2}, \nonumber\\
	& = \text{vol}_{d_s} (z+d_s-2)\mu e^{2(z-1)\phi_1} r_+^{z+d_s-2} = \text{vol}_{d_s} J_2^t,
\end{align}
which lines up exactly with our expectations.
From here, it is possible to compute the system's internal energy,
\begin{align}
	\mathcal{E} & = W_N + T\mathcal{S} + \mu_2 \mathcal{Q}_2, \nonumber \\
	& = \text{vol}_{d_s} d_s M r_+^{z+d_s} = \text{vol}_{d_s} \mathbbm{T}^t_{\;t},
\end{align}
which again is self-consistent: the energy contained is proportional to the black hole's mass, and given by the $tt$-component of the stress-energy tensor.

Next, let us consider what happens under the variation of $\phi_1$. This, of course, is an effect that exists only for $z\neq 1$ and couples the variation of the Lifshitz $U(1)$ field and the dilaton. We find
\begin{align}
	\frac{\partial W_N}{\partial \phi_1} & = -\text{vol}_{d_s}2d_s(z-1)(M-1)r_+^{z+d_s} , \nonumber \\
	& = \mu_1 \frac{\partial \mathcal{Q}_1}{\partial \phi_1}+\text{vol}_{d_s}\lambda_\phi \mathbbm{O}_\phi,
\end{align}
where
\begin{align}
	\mathcal{Q}_1 & = \text{vol}_{d_s}\sqrt{2(z-1)(z+d_s)}e^{-d_s\phi_1} = \text{vol}_{d_s}J_1^t.
\end{align}
As expected, the explicit response of both components is manifest and they are not independent of each other in the context of the static solution.
Finally, we compute the pressure $\mathpzc{p}$. Our system's trivial volume dependence means the pressure is just the negated thermodynamical potential density in the grand canonical ensemble --- namely the density of the ``Dirichlet" free energy $W_D = W_N - \mu_1 \mathcal{Q}_1$ --- and is given by
\begin{align}
	\mathpzc{p} & = -\frac{W_D}{\text{vol}_{d_s}}, \nonumber \\
	& = \left( M - \frac{1}{z-d_s}\frac{k^2 e^{-2(z-1)\phi_1}}{r_+^{2z}} \right)r_+^{z+d_s} = -\mathbbm{T}^{x^1}_{\;\;x^1}+\frac{k \mathcal{O}_k}{\text{vol}_{d_s}}.
\end{align}
Here we have introduced an operator dual to the impurity $k$,
\begin{align}
	\mathcal{O}_k \equiv -\frac{1}{d_s}\frac{\partial W_N}{\partial k} = -\text{vol}_{d_s}\frac{k e^{-2(z-1)\phi_1}r_+^{-z+d_s}}{z - d_s},
\end{align}
which functions analogously to a magnetization in response to an applied magnetic field \cite{Andrade2014}. The $1/d_s$ factor is chosen to normalize the response to one spatial coordinate. As expected, turning on impurities creates the disparity $\mathpzc{p} \neq - \mathbbm{T}^{x^1}_{\;\;x^1}$. The simple form of the pressure guarantees the satisfaction of a Smarr-like relation,
\begin{align}
	\epsilon + \mathpzc{p} = T s + \sum_{q=1}^2 \mu_q \rho_q,
\end{align}
where $\epsilon$, $s$, and $\rho_q$ are the energy, entropy and charge densities respectively.

\section{DC Conductivities}

When we consider fluctuations of the bulk spacetime and fields at the level of slowly-varying gradients, the equations of motion decouple into three separate modes. A so-called sound mode from which susceptibilities can be derived, a tensor mode which expresses vorticity of the fluid, and a vector mode which contains the system's heat and charge currents \cite{Kiritsis2015,Kiritsis2017}. To acquire the DC conductivities, we can supply linear time sources for the vector mode and extract the current responses. We take the ansatz,
\begin{subequations}
\begin{align}
	\delta ds^2 & = 2 r^2 \delta g_{rx^1}dr dx^1+2\left( -\frac{\nabla T}{T} r^{2z}f t+r^2 \delta g_{tx^1}\right)dt dx^1, \\
	\delta A_q & = \left( -\nabla \mu_q t+\frac{\nabla T}{T} A_{q,t} t + \delta A_{q,x^1} \right)dx^1, \\
	\delta \chi_I & = \delta_{I1}\delta \chi_1,
\end{align}
\end{subequations}
where the perturbations --- without loss of generality due to rotational symmetry --- are sourced along the $x^1$-direction. The equations of motion come in two batches,
\begin{widetext}
\begin{subequations}
\begin{align}
	& -f r^{z-d_s+1}\left( r^{-z+d_s+3}\delta g_{tx^1}'+\sum_{q=1}^2 \rho_q \delta A_{q,x^1}\right)' + k^2 X(\phi) \delta g_{tx^1} = 0, \\
	& j^{(\rho_q)\prime} = 0, \qquad  j^{(\rho_q)} = - r^{z+d_s-1}f Z_q(\phi) \delta A_{q,x^1}' - \rho_q \delta g_{tx^1},
\end{align}
\end{subequations}
\end{widetext}
which are $E_{tx^1}$ and $M_q^{x^1}$, and
\begin{widetext}
\begin{subequations}
\begin{align}
	& \left[r_+^{z+d_s+1}f'(r_+)-\frac{1}{z-d_s} k^2 X(\phi)r^{2(z-1)}\left( r^{-z+d_s} - r_+^{-z+d_s}\right)\right]\frac{\nabla T}{T}  \nonumber \\
	&\qquad\quad+\sum_{q=1}^2 \rho_q \nabla \mu_q - k X(\phi) r^{z+d_s+1}f (\delta\chi_1 ' - k \delta g_{rx^1}) = 0, \\
	& - k \frac{\nabla T}{T} + r^{z-d_s+1}\left[r^{-z+d_s+3} f\left(\delta\chi_1 ' - k \delta g_{rx^1}\right)\right]' = 0,
\end{align}
\end{subequations}
\end{widetext}
which are $E_{rx^1}$ and $\Xi_1$, respectively. We can see $\Xi_1 =0$ follows from $E_{tx^1}=0$. Thus, $E_{rx^1}$ completely decouples from the other equations and acts as a first-order constraint.

The electric currents $j^{(\rho_q)} = J_q^{x^1}$ are conserved in the bulk, but we can construct another conserved current by considering a Killing vector as shown in Appendix \ref{adxKilling}. The result is
\begin{align}
\begin{split}
	j^{(\mathpzc{q})} &= r^{3z+d_s-1} f^2 \left( r^{-2(z-1)}f^{-1} \delta g_{tx^1}\right) ' -\sum_{q=1}^2 A_{q,t} j^{(\rho_q)}\\
	j^{(\mathpzc{q})\prime} &= 0,
\end{split}
\end{align}
where and the conserved bulk quantity $j^{(\mathpzc{q})}$ is the boundary heat current.
We demand these functions are regular at the horizon in in-going Eddington-Finkelstein coordinates, given by the transformation
\begin{align}
	dt_+ = dt + \frac{dr}{r^{z+1}f},
\end{align}
and thus near the horizon we obtain $t = t_+ - \frac{1}{4\pi T} \ln (r-r_+)$. Regularity yields the asymptotic relations
\begin{subequations}
\begin{align}
	& \delta g_{rx^1} \sim \frac{1}{k^2 X(\phi_+) s T(r-r_+)} \left( s \nabla T + \sum_{q=1}\rho_q \nabla \mu_q \right), \\
	& \delta g_{tx^1} \sim \frac{4\pi}{k^2 X(\phi_+)s} \left( s \nabla T + \sum_{q=1}\rho_q \nabla \mu_q \right), \\
	& \delta A_{q,x^1} \sim \frac{1}{4\pi T}\nabla \mu_q\ln(r-r_+),
\end{align}
\end{subequations}
where we express our quantities in terms of the entropy and charge densities and defined $\phi_+ = \phi(r_+)$. Plugging into our currents, which are conserved in the bulk, we find
\begin{widetext}
\begin{subequations} \label{currents}
\begin{align}
	j^{(\mathpzc{q})} & = -\frac{4\pi T}{k^2 X(\phi_+)} \left( s \nabla T + \sum_{q=1}^2 \rho_q \nabla \mu_q\right), \\
	j^{(\rho_q)} & = -r_+^{d_s-2} Z_q(\phi_+) \nabla \mu_q - \frac{ 4\pi\rho_q}{k^2 X(\phi_+)s} \left( s \nabla T + \sum_{p=1}^2 \rho_p \nabla \mu_p \right).
\end{align}
\end{subequations}
\end{widetext}
The heat current is related to the energy current via
\begin{align}
	j^{(\mathpzc{q})} = j^{(\epsilon)} - \sum_{q=1}^2 \mu_q j^{(\rho_q)},
\end{align}
which of course is a measure of energy flow in excess of the energy due to charge transfer. Thus, the energy current is given by
\begin{align}
	j^{(\epsilon)} =& - r_+^{d_s-2} \sum_{q=1}^2 Z_q(\phi_+) \mu_q \nabla \mu_q \nonumber\\ 
	&-  \frac{4\pi \left(sT +\sum_{q=1}^2 \mu_q \rho_q\right)}{k^2 X(\phi_+) s} \left( s\nabla T + \sum_{p=1}^2 \rho_p \nabla \mu_p \right).
\end{align}

\section{Energy and Charge Diffusion}

Let us consider the diffusion of energy and charge in our system. From hereon we can simply work with densities. The energy and charges follow continuity equations,
\begin{align}\label{continuity}
	\partial_t \epsilon + \nabla\cdot j^{(\epsilon)} = 0, \quad \partial_t \rho_q + \nabla\cdot j^{(\rho_q)} = 0.
\end{align}
In the Neumann ensemble, gradients of $T$, $\phi_1$ and $\mu_2$ source gradients of energy and charge density. We will examine the diffusion of the system's energy and electric charge under the constraint where the ``Lifshitz charge" is completely fixed and uniform; that is, $j^{(\rho_1)} = 0$ and $\nabla \rho_1 = 0$. Under this constraint, 
\begin{subequations}
\label{sus}
\begin{align}
	& \nabla \epsilon = \left( c_{\mu_2} + \mu_2 \zeta \right) \nabla T + \left( T \zeta + \mu_2 \chi \right) \nabla \mu_2, \\
	& \nabla \rho_2 = \zeta \nabla T + \chi \nabla \mu_2,
\end{align}
\end{subequations}
where
\begin{subequations}
\begin{align}
	& c_{\mu_2} = T \frac{\partial s}{\partial T}\bigg|_{\rho_1,\mu_2} = - T \frac{\partial^2 w_N}{\partial T^2}, \\
	& \zeta = \frac{\partial s}{\partial \mu_2}\bigg|_{T,\rho_1} = \frac{\partial \rho_2}{\partial T}\bigg|_{\rho_1,\mu_2} = - \frac{\partial^2 w_N}{\partial T \partial \mu_2}, \\
	& \chi = \frac{\partial \rho_2}{\partial \mu_2}\bigg|_{T,\rho_1} = - \frac{\partial^2 w_N}{\partial \mu_2^2},
\end{align}
\end{subequations}
the susceptibilities are computable as second-order derivatives of the free energy density $w_N$.

The associated heat and charge currents,
\begin{align}\label{jconduct}
	& j^{(\mathpzc{q})} = -\overline{\kappa} \nabla T - T \alpha \nabla \mu_2, \qquad j^{(\rho_2)}  = -\alpha \nabla T - \sigma \nabla\mu_2, \nonumber\\
	& j^{(\epsilon)} = -\left(\overline{\kappa} + \mu_2 \alpha\right) \nabla T - \left( T \alpha + \mu_2 \sigma\right)\nabla \mu_2,
\end{align}
are given by, utilizing Eq. \eqref{currents} and the constraint $j^{(\rho_1)} = 0$,
\begin{subequations}
\begin{align}
	& \overline{\kappa} = \frac{4\pi s T}{\Sigma_1 k^2 X(\phi_+)}, \\
	& \alpha = \frac{4 \pi \rho_2}{\Sigma_1 k^2 X(\phi_+)}, \\
	& \sigma = r_+^{d_s-2} Z_2 (\phi_+) + \frac{4\pi\rho_2^2}{\Sigma_1 k^2 X(\phi_+)s}, \\
	& \Sigma_1 = 1+\frac{\rho_1^2}{k^2 X(\phi_+) Z_1(\phi_+) r_+^{2d_s-2}},
\end{align}
\end{subequations}
where $\Sigma_1$ measures the response due to application of $\nabla \mu_1$ on the non-Lifshitz matter. The application of this gradient is what allows the conductivities to be finite even in the absence of momentum dissipation, \textit{i.e.} $k \to 0$. This is an expected feature in a system with two species of $U(1)$ fields, first observed by Sonner \cite{Sonner2013} and later by Cremonini and Pope \cite{Cremonini2017}. Our conductivities calculated through linear time sources are, of course, identical to the Neumann conductivities found by Cremonini, Cveti\v{c} and Papadimitriou \cite{Cremonini2018}. This conductivity feature is an instance of some of the more robust behavior a $U(1) \times U(1)$ model can afford. 

The continuity equation \eqref{continuity} in concert with the conductivities \eqref{jconduct} and susceptibilities \eqref{sus} yields a diffusion equation for energy and charge:
\begin{align}
	\begin{pmatrix} \partial_t \rho_2 \\ \partial_t \epsilon \end{pmatrix} = \bm{D} \begin{pmatrix} \nabla^2 \rho_2 \\ \nabla^2 \epsilon  \end{pmatrix},
\end{align}
where the diffusion matrix $\bm{D}$ is determined from the conductivity matrix $\bm{\sigma}$ and susceptibility matrix $\bm{\chi}$ via the celebrated Einstein relation,
\begin{align}
	\bm{D} = \bm{\sigma}\bm{\chi}^{-1}.
\end{align}
The diffusion eigenvalues follow
\begin{subequations}
\begin{align}
	& D_+ D_- = \frac{\kappa}{c_{\rho_2}} \frac{\sigma}{\chi}, \\
	& D_+ + D_- = \frac{\kappa}{c_{\rho_2}} + \frac{\sigma}{\chi} + \frac{T \sigma}{c_{\rho_2}} \left(\frac{\zeta}{\chi} - \frac{\alpha}{\sigma} \right)^2,
\end{align}
\end{subequations}
where we define
\begin{align}
	c_{\rho_2} = c_{\mu_2} - \frac{T \zeta^2}{\chi}
\end{align}
as the specific heat for fixed electric charge --- which follows from Maxwell relations --- and
\begin{align}
	\kappa & = \overline{\kappa} - \frac{T \alpha^2}{\sigma}, \nonumber \\
	& = \frac{4\pi s T}{\Sigma_1 k^2 X(\phi_+)+\frac{4\pi \rho_2^2}{r_+^{d_s - 2}Z_2(\phi_+)s}}
\end{align}
to be the open-circuit thermal conductivity,  where no electric charge can flow. We note that $\kappa$ is explicitly dependent on both the metric and $A_1$, in contrast to traditional holographic systems where $\kappa$ is explicitly dependent only upon the form of the metric \cite{Blake2017}. This is a direct consequence of our fixed Lifshitz charge scenario.

\begin{figure}[t!]
	\centering
	\includegraphics[scale=1]{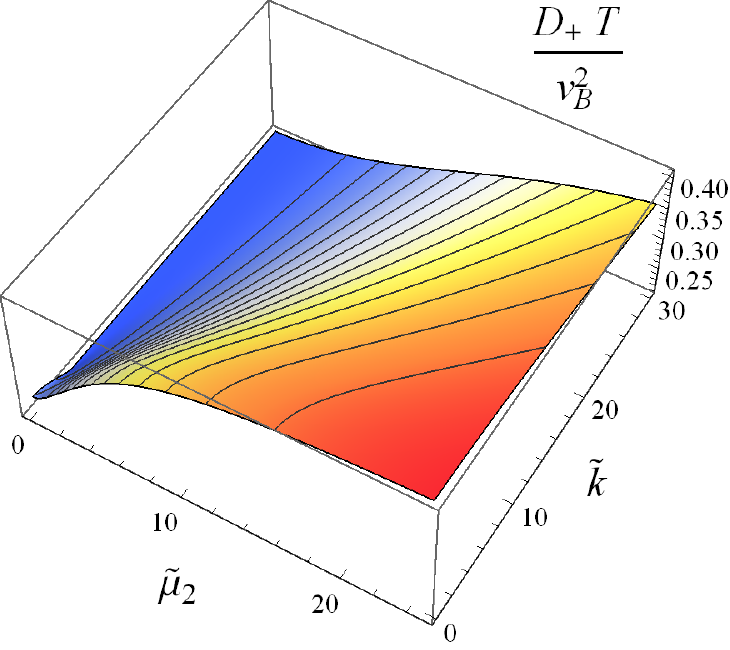}
	\includegraphics[scale=1]{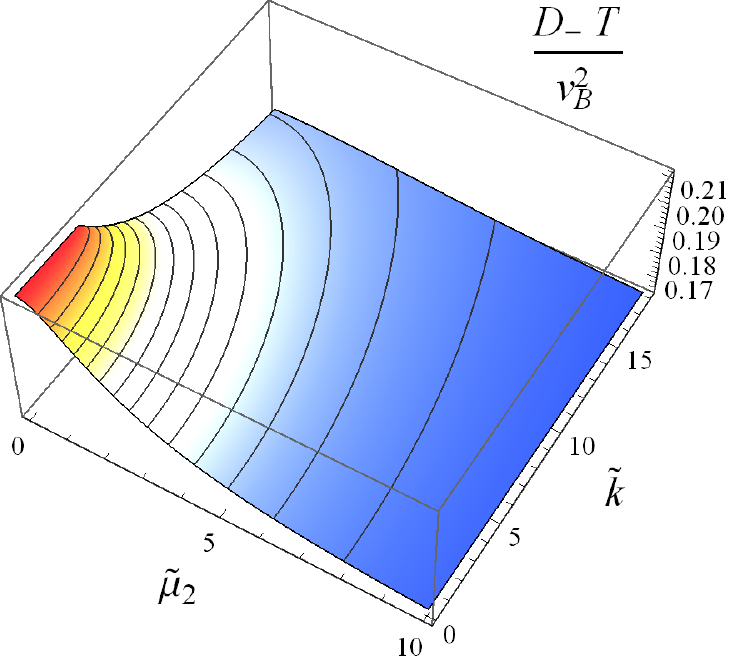}
	\captionsetup{justification=raggedright,singlelinecheck=false}
	\caption{$D_{\pm}T/v_B^2$ for $d_s = 2$ and $z=3/2$.}
	\label{smallz}
\end{figure}

\begin{figure}[t!]
	\centering
	\includegraphics[scale=1]{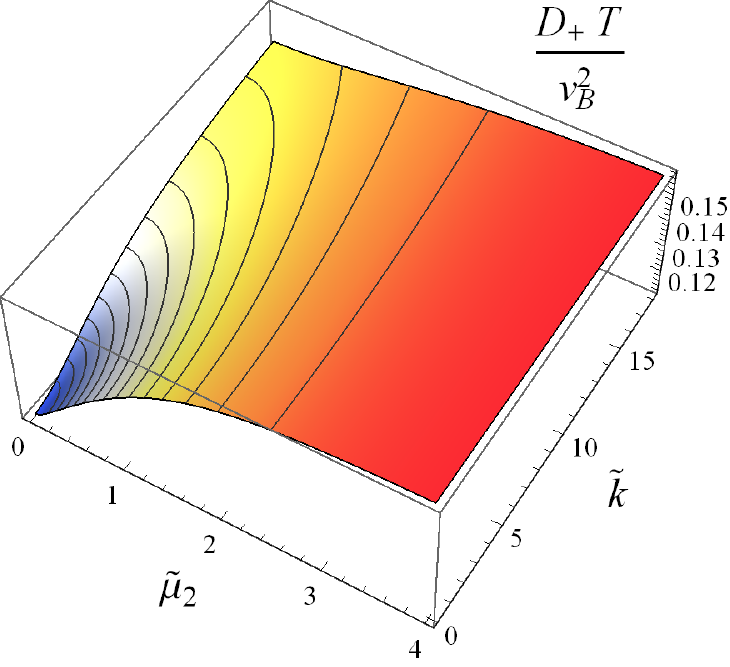}
	\includegraphics[scale=1]{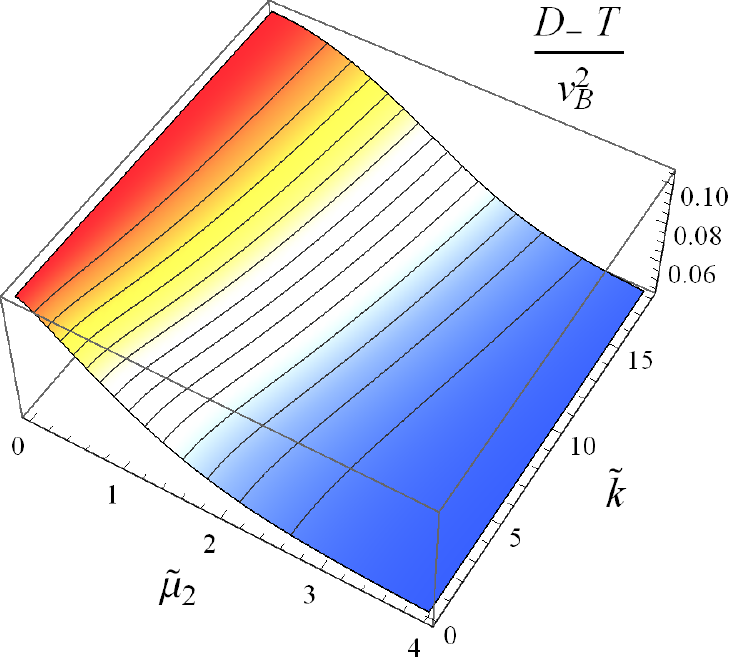}
	\captionsetup{justification=raggedright,singlelinecheck=false}
	\caption{$D_{\pm}T/v_B^2$ for $d_s = 3$ and $z=7/2$.}
	\label{bigzdiff}
\end{figure}

\begin{figure}[t!]
	\centering
	\includegraphics[scale=1]{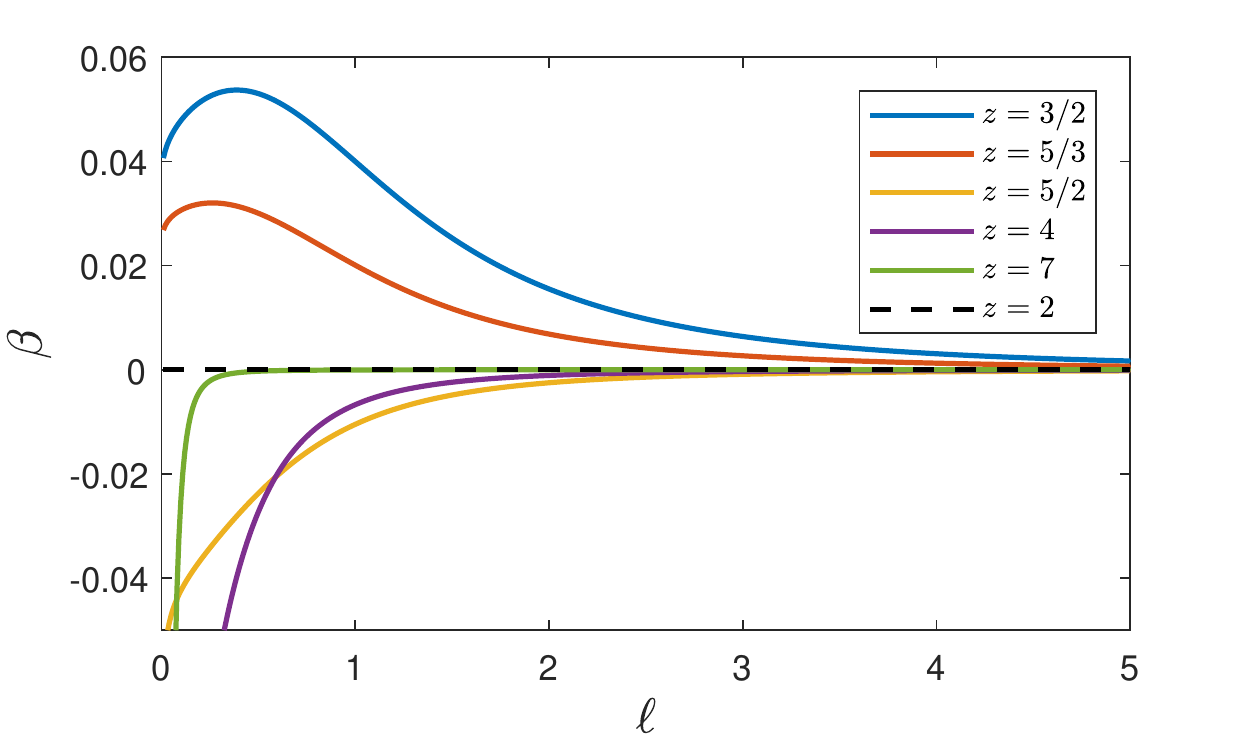}
	\captionsetup{justification=raggedright,singlelinecheck=false}
	\caption{Plot of the $\beta$-function (Eq. (\ref{betaf})) illustrating the universality of the diffusivities as a function of length for varying $z$, using $\tilde{\mu}_2 = 1.8$ and $d_s=3$. The qualitative features of $\beta$ are independent of the chemical potential and the number of spatial dimensions. The universal sign change signifies that the length dependence of the diffusivities is controlled by a fixed point at $z=2$.}
	\label{fig:beta}
\end{figure}

We will measure the diffusion eigenvalues relative to the butterfly velocity,
\begin{align}
	v_B^2 = \frac{2\pi}{d_s}T r_+^{z-2},
\label{butterfly}
\end{align}
which will make $D_{\pm}T/v_B^2$ pure numbers. This velocity is proposed to be the characteristic velocity for Lifshitz geometries and is independent of all matter content \cite{Blake2016}. Rescaling Eq. \eqref{temp} yields 
\begin{align}
	& 1 = \frac{R^z}{4\pi}\left( z+ d_s - \frac{\left(z+d_s-2\right)^2}{2d_s}\tilde{\mu}_2^2 R^{-2} - \frac{1}{2} \tilde{k}^2 R^{-2z}\right), \\
	& r_+ = T^{1/z} R\left(\tilde{\mu}_2, \tilde{k}\right), \quad \tilde{\mu}_2 = \frac{\mu_2 e^{(z-1)\phi_1}}{T}, \quad \tilde{k} = \frac{k e^{-(z-1)\phi_1}}{T^z},
\end{align}
and all pure numbers in our system will be functions of the two parameters $\tilde{\mu}_2$ and $\tilde{k}$. Written in scaling form, the determinant and trace of the diffusion matrix are given, respectively, by
\begin{widetext}
\begin{subequations}
\begin{align}
	& \frac{D_+ D_- T^2}{v_B^4} = \frac{z d_s \left[2(z+d_s)+\tilde{k}^2 R^{-2z}\right] - (z-2)(z+d_s-2)^2 \tilde{\mu}^2 R^{-2} }{8\pi^2(z+d_s-2)\left[2(z-1)(z+d_s) + \tilde{k}^2 R^{-2z} \right]}, \\
	& \frac{(D_+ + D_-)T}{v_B^2}= \frac{d_s}{2\pi(z+d_s-2)}+\frac{z d_s \left[2(z+d_s)+\tilde{k}^2 R^{-2z}\right]+(z-2)\left[(z-2)^2 - d_s^2\right]\tilde{\mu}_2^2 R^{-2}}{4\pi d_s\left[2(z-1)(z+d_s) + \tilde{k}^2 R^{-2z} \right]}.
\end{align}
\end{subequations}
\end{widetext}
We will always order the eigenvalues such that $D_+ \geq D_-$. We display a couple of solutions explicitly in Figs. \ref{smallz} and \ref{bigzdiff}. Notably, we find the diffusion eigenvalues are bounded. The bounds can be obtained analytically through various limits of $\tilde{\mu}_2$ and $\tilde{k}$ and are given by
\begin{figure}[t!]
	\centering
	\includegraphics[scale=1]{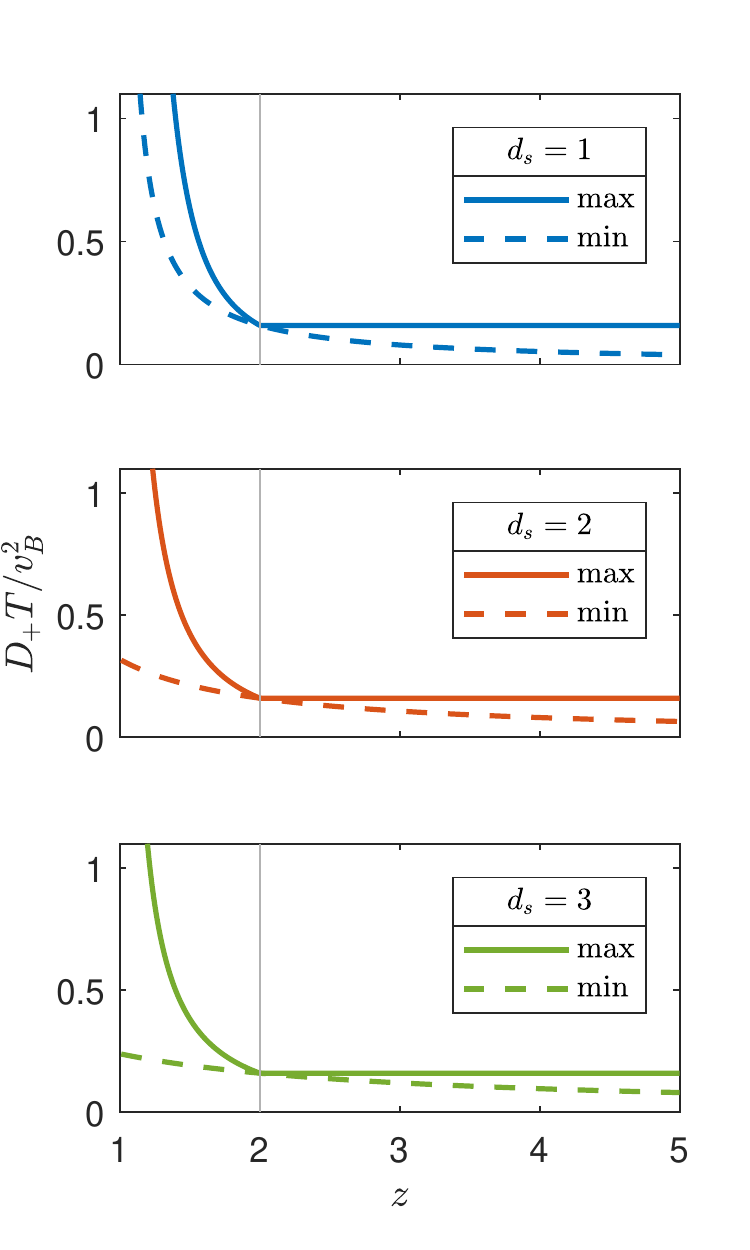}
	\includegraphics[scale=1]{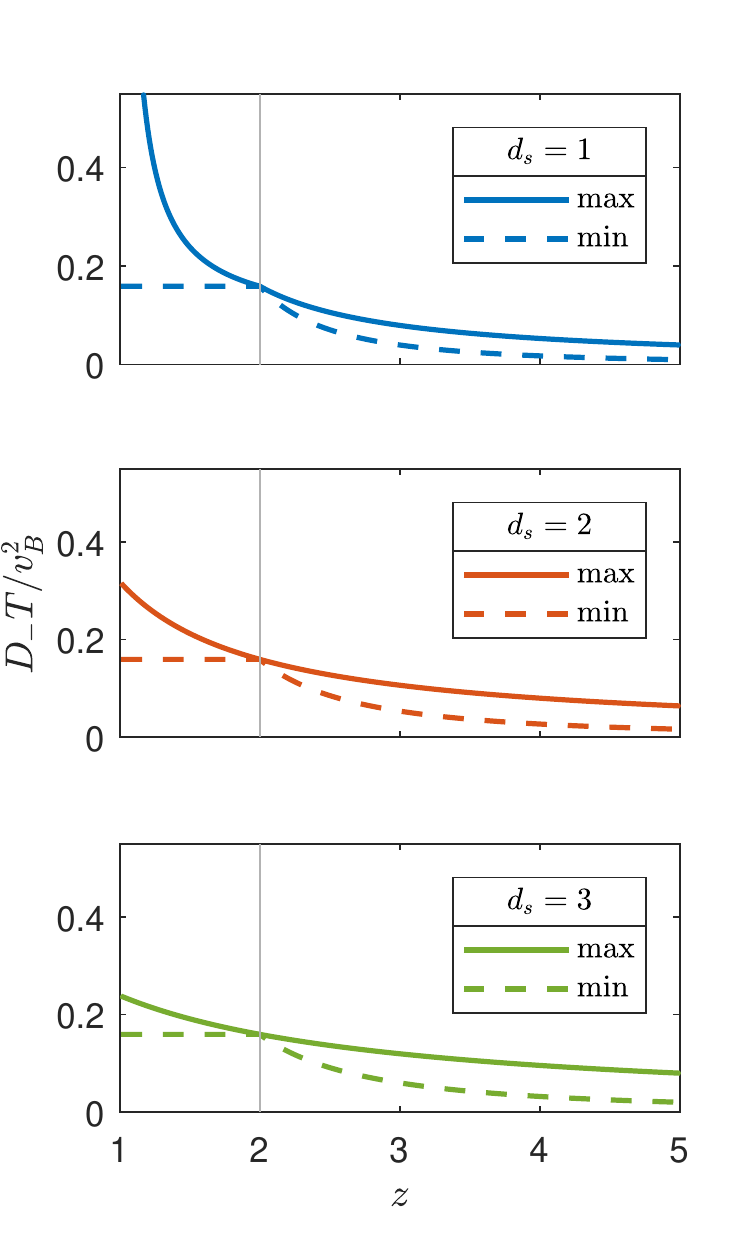}
	\captionsetup{justification=raggedright,singlelinecheck=false}
	\caption{The bounds of $D_{\pm}T/v_B^2$ for $d_s = \{1,2,3\}$. The minimum of $D_+ T/v_B^2$ asymptotes to $1/4\pi$ as $z\to \infty$.}
	\label{difbounds}
\end{figure}
\begin{subequations}
\begin{align}
	& \max\left(\frac{D_+ T}{v_B^2}\right) = \begin{cases} \frac{d_s}{2\pi(z-1)(z+d_s-2)} & z < 2  \\  \frac{1}{2\pi} & z\geq 2\end{cases}, \\
	& \min \left(\frac{D_+ T}{v_B^2}\right) =  \begin{cases} \frac{z}{4\pi(z-1)} & z > 2 \text{ and } z > d_s  \\  \frac{d_s}{2\pi(z+d_s-2)} & \text{else}\end{cases}, \\
	& \max\left(\frac{D_- T}{v_B^2}\right) = \begin{cases} \frac{z}{4\pi(z-1)} & d_s < z < 2  \\  \frac{d_s}{2\pi(z+d_s-2)} & \text{else}\end{cases}, \\
	& \min \left(\frac{D_- T}{v_B^2}\right) = \begin{cases} \frac{1}{2\pi} & z \leq 2  \\ \frac{d_s}{2\pi(z-1)(z+d_s-2)}  & z > 2\end{cases},
\end{align}
\end{subequations}
which are shown in Fig. \ref{difbounds} for $d_s = \{1,2,3\}$, in which the primacy of $z=2$  as the scale-invariant point of the diffusion equation is evident.  To illustrate this further, we plot Eq. (\ref{betaf}) with the length defined as $\ell=2\pi/{\tilde{k}}$ in Fig. \ref{fig:beta}.\footnote{We could very well have used $\det \bm{D}$ instead of $\text{tr}\; \bm{D}$, though the qualitative features are identical.}  This characteristic length functions as an effective lattice spacing.  As is evident, there is a universal sign change at $z=2$.  At $z=2$, the diffusion constants are equal and given by the universal value,
\begin{align}
	 D_{\pm}= \frac{1}{d_s},
\end{align}
 as remarked in the introduction. The universal nature of the charge and energy diffusion constants stems from the underlying scale invariance of the diffusion equation when $z=2$.  The sign change of $\beta$ signals a fixed point exists at $z=2$ analogous to the scale-invariance that obtains for the conductance in the case of $d_s=2$ in Anderson localization \cite{Anderson1979}.   Necessarily, the diffusion scale $v_B^2/T$ is also a pure number at this point, as seen in Eq. \eqref{butterfly} where the horizon dependence vanishes. The marginality of $z=2$ has been noted in other contexts such as a Lifshitz string \cite{Tong2013} and the stability of scalar hair \cite{Mozaffar2013}. In particular, the so-called perfect fluid, which depends only upon its own rest mass and isotropic pressure, can only exist in concert with Galilean boosts at $z=2$ wherein the scale-dependent mass contribution drops out \cite{deBoer2018}. Note that we have allowed non-integer values of $d_s$ in our solutions, which can effectively be obtained through the use of a hyperscaling violating parameter. 


To put our results for the transport in the context of expected results, we first observe the universal features. In the decoupling limit where the chemical potential is turned off, we find the charge and energy diffusion constants follow
\beq
	\lim_{\tilde{\mu}_2\to 0} \frac{\sigma}{\chi}\frac{T}{v_B^2} &=& \frac{d_s}{2\pi(z+d_s-2)},\nonumber\\
	\lim_{\tilde{\mu}_2,\tilde{k} \to 0}\frac{\kappa}{c_{\rho_2}}\frac{T}{v_B^2} &=& \frac{z}{4\pi(z-1)},
	\label{eq:diffdecouple}
\eeq
which exactly match the purported decoupled forms in \cite{Blake2016, Blake2017}.
For $z \geq 2$ and $z \geq d_s$, which heads toward the decoupling limit of large $z$, we find these forms match $\max(D_- T/v_B^2)$ and $\min(D_+ T/ v_B^2)$ exactly. The limits in which deviations occur are laid bare in Fig. \ref{bigzdiff}. For $D_-$,  deviation from this behavior is found when $\tilde{\mu}_2 \gg 1$, the limit in which we expect thermoelectric interactions to be quite strong and so inhibit charge flow. For $D_+$, deviation is found under either the conditions $\tilde{k} \gg 1$ or $\tilde{\mu}_2 \gg 1$, emphasizing that energy diffusivity is heavily subject to all matter interactions.


A limit of interest is $z \to \infty$. In particular, the bounds indicate clearly that $D_- T/v_B^2 \to 0$, signaling that charge does not diffuse in this limit, only energy. Conventionally, $z \to \infty$ corresponds to localized critical physics; the divergence of the critical length guarantees no dynamic critical behavior can obtain on any appreciable time scale. For our system, we can interpret this to mean charge must follow this type of quantum critical behavior but energy does not. We also notice the saturation of the $D_+$ diffusion constant for $z \geq 2$, whereupon $ 1/4\pi \leq D_+ T/v_B^2 \leq 1/2\pi$. The upper bound is the typical saturation observed in the SYK model \cite{Davison2017}. Intuitively, this characterizes the fact that energy diffusivity must be bounded from below and above, so long as momentum dissipation is present, regardless of the value of $z$.

\section{Conclusions}

From this detailed treatment of Lifshitz holography, we have been able to derive a series of thermodynamic and dynamical response functions. Of particular note is the explicit derivation of both the Bekenstein-Hawking and Smarr-like relationships, made possible by the exact computation of the renormalized thermodynamic potential.   Our calculations reveal the universal features of the diffusion constants near $z=2$ even in the complicated setting in which charge and thermal degrees of freedom are treated on equal footing.  This universality obtains because of the emergence of a fixed point  characterizing the length dependence of the diffusivities at $z=2$.   The vanishing of the charge diffusion constant in the local critical limit of $z\rightarrow\infty$ represents the ultimate deviation from the expected bounds.  
Since our treatment fully incorporates thermodynamics and electrical responses, it should serve as a template for extracting the coterie of transport coefficients relevant to quantum critical matter.

\appendix

\section{Full Background Renormalization}\label{backren}

For $k \neq 0$, we can have a possibly infinite number of divergent terms in Eq. \eqref{boundvary}. Consider counter-terms
\begin{widetext}
\begin{align}
	I_{\text{c.t.}} = \int_{\partial \mathcal{M}}\sum_{n=0}\left(c_{n}^{\text{(vol})}\sqrt{-\gamma}+c_n^{(J_1^2)}\frac{J_1^2}{\sqrt{-\gamma}Z_1(\phi)}\right)\left[X(\phi)\sum_{I=1}^{d_s}\left(\hat{\nabla} \chi_I\right)^2\right]^n,
\end{align}
\end{widetext}
whose contributions to our boundary responses are
\begin{widetext}
\begin{align}
	& \mathbbm{T}^a_{\;\;b|\text{c.t.}} = \sum_{n=0}\left[ X(\phi)\sum_{J=1}^{d_s} \left(\hat{\nabla}\chi_J\right)^2\right]^{n-1}2X(\phi)\sum_{I=1}^{d_s}\Bigg[c_n^{(\text{vol})} \sqrt{-\gamma} \left(\frac{1}{2}\delta^a_{\;\;b}\left(\hat{\nabla}\chi_I\right)^2 - n \hat{\nabla}^a \chi_I \hat{\nabla}_b \chi_I\right) \nonumber \\
	& \qquad\qquad\quad +\frac{c_n^{(J_1^2)}}{\sqrt{-\gamma}Z_1(\phi)}\left( J_1^a J_{1,b} \left(\hat{\nabla}\chi_I\right)^2 - n J_1^2 \hat{\nabla}^a\chi_I \hat{\nabla}_b \chi_I -\frac{1}{2}\delta^a_{\;\;b} J_1^2 \right) \left(\hat{\nabla}\chi_I\right)^2\Bigg], \\
	& \mathbbm{A}_{1,a|\text{c.t.}} = \sum_{n=0}\left[X(\phi)\sum_{I=1}^{d_s}  \left(\hat{\nabla}\chi_I\right)^2\right]^{n}\frac{2 c_n^{(J_1^2)} J_{1,a}}{\sqrt{-\gamma}Z_1(\phi)}, \\
	& \mathbbm{O}_{\phi|\text{c.t.}} = \sum_{n=0}\left[X(\phi)\sum_{I=1}^{d_s}  \left(\hat{\nabla}\chi_I\right)^2\right]^{n}\nonumber\\
	&\qquad\qquad\quad\times\bigg[c_n^{(\text{vol})}\sqrt{-\gamma}\frac{n X'(\phi)}{X(\phi)}+c_n^{(J_1^2)} \frac{J_1^2}{\sqrt{-\gamma}Z_1(\phi)} \left( \frac{nX'(\phi)}{X(\phi)} - \frac{Z_1'(\phi)}{Z_1(\phi)}\right)\bigg].
\end{align}
\end{widetext}
The covariant derivative $\hat{\nabla}_a$ is defined from the intrinsic metric $\gamma_{ab}$. In general, we can increment $n$ to cancel out higher order powers involving $k^2$ terms that bleed over from our established action. The constants are given by
\begin{align}
	& c_{n}^{(\text{vol})} = \vartheta_n^{(\text{vol})}\frac{(2n-1)(z-1)-2d_s}{\left[4d_s(d_s-z)\right]^n}, \\
	& c_n^{(J_1^2)} = \vartheta_n^{(J_1^2)}\frac{1}{(z+d_s)\left[4d_s(d_s-z)\right]^n},
\end{align}
and are determined recursively from $n=0$ with $\vartheta_n^{(\text{vol})} = \{ -1, 1, \frac{1}{2}, \frac{1}{2}, \frac{5}{8}, \frac{7}{8}, \dots  \}$ and $\vartheta_n^{(J_1^2)} = \{\frac{1}{2}, \frac{1}{2}, \frac{3}{4}, \frac{5}{4}, \frac{35}{16}, \frac{63}{16}, \dots  \}$. This is a specific encoding of the renormalization provided by the Hamilton-Jacobi equations, which expands in a series of functional derivatives \cite{Papadimitriou2011,Chemissany2015}.

\section{Killing Vector Conserved Quantity}\label{adxKilling}

Here we determine the bulk conserved quantity that will be dual to the heat current as outlined in \cite{Donos2014}. Suppose there exists a Killing vector $\xi$, defined by
\begin{align}
	\mathcal{L}_\xi g_{ab} = \nabla_{(a}\xi_{b)} = 0,
\end{align}
which of course can correspond to an infinitesimal diffeomorphism. We consider that the Lie derivatives on our physical observable fields vanishes; that is
\begin{align}
	\mathcal{L}_\xi F_{q,ab} = \mathcal{L}_\xi \phi = \mathcal{L}_\xi \chi_I = 0.
\end{align}
The first of these, rewritten, states that
\begin{align}
	\left(i_\xi d + d i_\xi\right) F_q = 0,
\end{align}
and thus we can assume that $i_\xi F_q$ is an exact form
\begin{align}
	i_\xi F_q = d\theta_q
\end{align}
for some functions $\theta_q$. This also implies that we can express
\begin{align}
	\mathcal{L}_\xi A_q = d\psi_q
\end{align}
for some functions $\psi_q$. These identities will allow us to construct a total derivative by examining
\begin{widetext}
\begin{align}
	\nabla_b \nabla^a \xi^b & = R^a_{\;\;b} \xi^b, \nonumber \\
	& = \frac{1}{d_s}V(\phi) \xi^a + \frac{1}{2}\sum_{q=1}^2\left(\xi^c Z_q(\phi) F_{q}^{ab}F_{q,cb}-\frac{1}{2d_s}\xi^a Z_q(\phi)F_q^2\right).
\end{align}
\end{widetext}
This expression can be rearranged as
\begin{align}
	\nabla_b G^{ab} = \frac{1}{d_s}V(\phi)\xi^a,
\end{align}
where
\begin{align}
	G^{ab} = \nabla^a \xi^b + \frac{1}{2d_s}\sum_{q=1}^2 Z_q(\phi)\left[\left(\psi_q - d_s \theta_q\right)F_q^{ab} + 2\xi^{[a} F_q^{b]c}A_{q,c}\right].
\end{align}
To deduce this expression, we use the identities
\begin{align}
	& \xi^c Z_q(\phi) F_q^{ab}F_{q,cb} = \nabla_b\left( \theta_q Z_q(\phi) F_q^{ab} \right), \\
	& \xi^a Z_q(\phi) F_q^2 = \nabla_b\left( 4\xi^{[a} Z_q(\phi) F_q^{b]c}A_{q,c} + 2 \psi_q Z_q(\phi) F_q^{ab}\right),
\end{align}
where the latter of these can be realized from rearranging the Lie derivative $\mathcal{L}_\xi F_{q,ab} = \xi^c \nabla_c F_{q,ab} + F_{q,bc}\nabla_a \xi^c + F_{q,ac} \nabla_b\xi^c=0$. Now, as long as any components of $\xi$ vanish we can deduce a conserved quantity. By choosing $\xi = \nabla_t$, it is clear that the $x^i$ components then generate conserved quantities, which are dual to the boundary heat currents.

\section{Renormalization for Currents}\label{adxRenorm}

With the counter-terms provided in Eq. \eqref{counterterms}, we find that
\begin{align}
	\mathbbm{T}^{x^1}_{\;\;t} = j^{(\epsilon)}
\end{align}
and as such we would like to construct our boundary theory such that this is the response to our metric variation. Presently, we would find that the boundary variation leaves a source term \cite{Donos2014} in the action. This is not a problem for $z = 1$, but otherwise this term is divergent. To mollify this divergence, we can recast the intrinsic metric $\gamma_{ab}$ as
\begin{align}
	\gamma_{ab}dx^a dx^b = -n^2\left( dt - \frac{n_i}{n^2}dx^i \right)^2 + \sigma_{ij}dx^i dx^j,
\end{align}
akin to the ADM formalism but where our boundary spacetime is instead foliated by the normalized time-like covector $\frac{1}{n}\nabla_t$. Then we can consider the fundamental variational objects of our theory to be $n$, $n_i$ and $\sigma_{ij}$ instead of the boundary metric. Our variation becomes
\begin{widetext}
\begin{align}
	\delta I = \int_{\partial \mathcal{M}} \left[\left( -\mathbbm{T}^{tt}n^2 + \mathbbm{T}^{ij}\frac{n_i n_j}{n^2}\right)\frac{\delta n}{n} -\mathbbm{T}^i_{\;t} \frac{\delta n_i}{n^2}+\frac{1}{2} \mathbbm{T}^{ij}\delta\sigma_{ij} + \cdots \right],
\end{align}
\end{widetext}
and now the source term present in the $\delta n_i$ term decays. Additionally, thanks to our formulation of $n_i$ as a small parameter, the response to the normalized variation of the lapse $\delta n/n$ is exactly $\mathbbm{T}^t_{\;t}$. Thus nothing about our static background scheme is modified.

\section{Linear Time Sources}

Linear time sources provide a straightforward scheme for deducing DC response functions. Additionally, through our gauge symmetries, they have a clear interpretation as temperature and chemical potential gradients. We can transform the metric and $U(1)$ fields as
\begin{align}
	& g_{ab} \to g_{ab} + \mathcal{L}_\xi g_{ab}, \\
	& A_{q,a} \to A_{q,a} + \mathcal{L}_\xi A_{q,a} + \nabla_a \Lambda_q,
\end{align}
where $\xi$ parameterizes an infinitesimal coordinate transformation $x^a \to x^a + \xi^a$ and each $\Lambda_q$ a $U(1)$ transformation, respectively. For the choices
\begin{align}
	& \xi = - t x^1 \frac{\nabla T}{T} \nabla_t, \\
	& \Lambda_q = t x^1 \nabla\mu_q,
\end{align}
our ansatz sources become
\begin{align}
	& \delta ds^2 = 2 r^{2z} f x^1\frac{\nabla T}{T}dt^2, \\
	& \delta A_q = x^1 \left(\nabla \mu_q - A_{q,t}\frac{\nabla T}{T}\right)dt,
\end{align}
which are exactly the gradients we would expect for small perturbations of $T$ and $\mu_q$. The form of the temperature fluctuations are determined by perturbing the period of Euclidean time, $1/T$.

\acknowledgments

We would like to thank Matteo Baggioli, Garrett Vanacore, Tom Faulkner, Rob Leigh, Michael Blake, Sean Hartnoll and Julian Sonner for comments and discussion along the way.  We also thank the NSF DMR-1461952 for partial funding of this project.

\bibliographystyle{jhep}

\providecommand{\href}[2]{#2}\begingroup\raggedright\endgroup


\begin{thebibliography}{10}

\bibitem{Bianconi2018}
A.~Bianconi, \emph{Lifshitz transitions in multi-band hubbard models for
  topological superconductivity in complex quantum matter},
  \href{https://doi.org/10.1007/s10948-017-4535-1}{\emph{Journal of
  Superconductivity and Novel Magnetism} {\bfseries 31} (2018) 603}.

\bibitem{Norman2010}
M.~R. Norman, J.~Lin and A.~J. Millis, \emph{Lifshitz transition in underdoped
  cuprates}, \href{https://doi.org/10.1103/PhysRevB.81.180513}{\emph{Phys. Rev.
  B} {\bfseries 81} (2010) 180513}.

\bibitem{Shi2018}
X.~Shi, Z.-Q. Han, X.-L. Peng, P.~Richard, T.~Qian, X.-X. Wu et~al.,
  \emph{Enhanced superconductivity accompanying a Lifshitz transition in
  electron-doped fese monolayer}, {\emph{Nature Communications} {\bfseries 8}
  (2017) 14988 EP }.

\bibitem{Volovik2018}
G.~E. Volovik, \emph{Topological Lifshitz transitions},
  \href{https://doi.org/10.1063/1.4974185}{\emph{Low Temperature Physics}
  {\bfseries 43} (2017) 47}
  [\href{https://arxiv.org/abs/https://doi.org/10.1063/1.4974185}{{\ttfamily
  https://doi.org/10.1063/1.4974185}}].

\bibitem{Kachru2008}
S.~Kachru, X.~Liu and M.~Mulligan, \emph{Gravity duals of Lifshitz-like fixed
  points}, \href{https://doi.org/10.1103/PhysRevD.78.106005}{\emph{Phys. Rev.
  D} {\bfseries 78} (2008) 106005}.

\bibitem{Horava2009}
P.~Ho\ifmmode~\check{r}\else \v{r}\fi{}ava, \emph{Quantum gravity at a Lifshitz
  point}, \href{https://doi.org/10.1103/PhysRevD.79.084008}{\emph{Phys. Rev. D}
  {\bfseries 79} (2009) 084008}.

\bibitem{Balasubramanian2011}
K.~Balasubramanian and J.~McGreevy, \emph{The particle number in Galilean
  holography}, \href{https://doi.org/10.1007/JHEP01(2011)137}{\emph{Journal of
  High Energy Physics} {\bfseries 2011} (2011) 137}.

\bibitem{Kiritsis2013}
E.~Kiritsis, \emph{Lorentz violation, gravity, dissipation and holography},
  \href{https://doi.org/10.1007/JHEP01(2013)030}{\emph{Journal of High Energy
  Physics} {\bfseries 2013} (2013) 30}.

\bibitem{Jensen2018}
K.~Jensen, \emph{On the coupling of Galilean-invariant field theories to curved
  spacetime}, \href{https://doi.org/10.21468/SciPostPhys.5.1.011}{\emph{SciPost
  Phys.} {\bfseries 5} (2018) 11}.

\bibitem{Gouteraux2011}
B.~Gout{\'e}raux and E.~Kiritsis, \emph{Generalized holographic quantum
  criticality at finite density},
  \href{https://doi.org/10.1007/JHEP12(2011)036}{\emph{Journal of High Energy
  Physics} {\bfseries 2011} (2011) 36}.

\bibitem{Fan2013}
Z.~Fan, \emph{Holographic superconductors with hyperscaling violation},
  \href{https://doi.org/10.1007/JHEP09(2013)048}{\emph{Journal of High Energy
  Physics} {\bfseries 2013} (2013) 48}.

\bibitem{Gouteraux2014}
B.~Gout{\'e}raux, \emph{Charge transport in holography with momentum
  dissipation}, \href{https://doi.org/10.1007/JHEP04(2014)181}{\emph{Journal of
  High Energy Physics} {\bfseries 2014} (2014) 181}.

\bibitem{Tarrio2012}
J.~Tarrio, \emph{Asymptotically Lifshitz black holes in
  Einstein-Maxwell-dilaton theories},
  \href{https://doi.org/10.1002/prop.201200011}{\emph{Fortschritte der Physik}
  {\bfseries 60} 1098}
  [\href{https://arxiv.org/abs/https://onlinelibrary.wiley.com/doi/pdf/10.1002/prop.201200011}{{\ttfamily
  https://onlinelibrary.wiley.com/doi/pdf/10.1002/prop.201200011}}].

\bibitem{Kiritsis2015}
E.~Kiritsis and Y.~Matsuo, \emph{Charge-hyperscaling violating Lifshitz
  hydrodynamics from black-holes},
  \href{https://doi.org/10.1007/JHEP12(2015)076}{\emph{Journal of High Energy
  Physics} {\bfseries 2015} (2015) 1}.

\bibitem{Kiritsis2017}
E.~Kiritsis and Y.~Matsuo, \emph{Hyperscaling-violating Lifshitz hydrodynamics
  from black-holes: part ii},
  \href{https://doi.org/10.1007/JHEP03(2017)041}{\emph{Journal of High Energy
  Physics} {\bfseries 2017} (2017) 41}.

\bibitem{Papadimitriou2014}
I.~Papadimitriou, \emph{Hyperscaling violating Lifshitz holography},
  \href{https://doi.org/https://doi.org/10.1016/j.nuclphysbps.2015.09.240}{\emph{Nuclear
  and Particle Physics Proceedings} {\bfseries 273-275} (2016) 1487 }.

\bibitem{Blake2016}
M.~Blake, \emph{Universal charge diffusion and the butterfly effect in
  holographic theories},
  \href{https://doi.org/10.1103/PhysRevLett.117.091601}{\emph{Phys. Rev. Lett.}
  {\bfseries 117} (2016) 091601}.

\bibitem{Blake2017}
M.~Blake, R.~A. Davison and S.~Sachdev, \emph{Thermal diffusivity and chaos in
  metals without quasiparticles},
  \href{https://doi.org/10.1103/PhysRevD.96.106008}{\emph{Phys. Rev. D}
  {\bfseries 96} (2017) 106008}.

\bibitem{Anderson1979}
E.~Abrahams, P.~W. Anderson, D.~C. Licciardello and T.~V. Ramakrishnan,
  \emph{Scaling theory of localization: Absence of quantum diffusion in two
  dimensions}, \href{https://doi.org/10.1103/PhysRevLett.42.673}{\emph{Phys.
  Rev. Lett.} {\bfseries 42} (1979) 673}.

\bibitem{Pang2010}
D.-W. Pang, \emph{Conductivity and diffusion constant in Lifshitz backgrounds},
  \href{https://doi.org/10.1007/JHEP01(2010)120}{\emph{Journal of High Energy
  Physics} {\bfseries 2010} (2010) 120}.

\bibitem{Ge2018}
X.-H. Ge, S.-J. Sin, Y.~Tian, S.-F. Wu and S.-Y. Wu, \emph{Charged BTZ-like
  black hole solutions and the diffusivity-butterfly velocity relation},
  \href{https://doi.org/10.1007/JHEP01(2018)068}{\emph{Journal of High Energy
  Physics} {\bfseries 2018} (2018) 68}.

\bibitem{Hartnoll2014}
S.~A. Hartnoll, \emph{Theory of universal incoherent metallic transport},
  \href{https://doi.org/10.1038/nphys3174}{\emph{Nature Physics} {\bfseries 11}
  (2014) 54}.

\bibitem{Hartman2017}
T.~Hartman, S.~A. Hartnoll and R.~Mahajan, \emph{Upper bound on diffusivity},
  \href{https://doi.org/10.1103/PhysRevLett.119.141601}{\emph{Phys. Rev. Lett.}
  {\bfseries 119} (2017) 141601}.

\bibitem{Cassani2011}
D.~Cassani and A.~F. Faedo, \emph{Constructing Lifshitz solutions from AdS},
  \href{https://doi.org/10.1007/JHEP05(2011)013}{\emph{Journal of High Energy
  Physics} {\bfseries 2011} (2011) 13}.

\bibitem{Andrade2013}
T.~Andrade and S.~F. Ross, \emph{Boundary conditions for metric fluctuations in
  Lifshitz}, {\emph{Classical and Quantum Gravity} {\bfseries 30} (2013)
  195017}.

\bibitem{Papadimitriou2016}
I.~Papadimitriou, \emph{Hyperscaling violating Lifshitz holography},
  \href{https://doi.org/https://doi.org/10.1016/j.nuclphysbps.2015.09.240}{\emph{Nuclear
  and Particle Physics Proceedings} {\bfseries 273-275} (2016) 1487 }.

\bibitem{Balasubramanian1999}
V.~Balasubramanian and P.~Kraus, \emph{A stress tensor for anti-de Sitter
  gravity}, \href{https://doi.org/10.1007/s002200050764}{\emph{Communications
  in Mathematical Physics} {\bfseries 208} (1999) 413}.

\bibitem{Chemissany2012}
W.~Chemissany, D.~Geissb{\"u}hler, J.~Hartong and B.~Rollier, \emph{Holographic
  renormalization for z = 2 Lifshitz spacetimes from AdS}, {\emph{Classical and
  Quantum Gravity} {\bfseries 29} (2012) 235017}.
 

\bibitem{Keranen2012}
V.~Ker{\"a}nen and L.~Thorlacius, \emph{Thermal correlators in holographic
  models with Lifshitz scaling}, {\emph{Classical and Quantum Gravity}
  {\bfseries 29} (2012) 194009}.

\bibitem{Taylor2015}
M.~Taylor, \emph{Lifshitz holography}, {\emph{Classical and Quantum Gravity}
  {\bfseries 33} (2016) 033001}.

\bibitem{Cremonini2017}
S.~Cremonini, H.-S. Liu, H.~L{\"u} and C.~Pope, \emph{DC conductivities from
  non-relativistic scaling geometries with momentum dissipation},
  \href{https://doi.org/10.1007/JHEP04(2017)009}{\emph{Journal of High Energy
  Physics} {\bfseries 2017} (2017) 9}.

\bibitem{Cremonini2018}
S.~Cremonini, M.~Cveti{\v{c}} and I.~Papadimitriou, \emph{Thermoelectric DC
  conductivities in hyperscaling violating Lifshitz theories},
  \href{https://doi.org/10.1007/JHEP04(2018)099}{\emph{Journal of High Energy
  Physics} {\bfseries 2018} (2018) 99}.

\bibitem{Alishahiha2012}
M.~Alishahiha, E.~{\'O}. Colg{\'a}in and H.~Yavartanoo, \emph{Charged black
  branes with hyperscaling violating factor},
  \href{https://doi.org/10.1007/JHEP11(2012)137}{\emph{Journal of High Energy
  Physics} {\bfseries 2012} (2012) 137}.

\bibitem{Kuang2017}
X.-M. Kuang and J.-P. Wu, \emph{Analytical shear viscosity in hyperscaling
  violating black brane},
  \href{https://doi.org/https://doi.org/10.1016/j.physletb.2017.08.060}{\emph{Physics
  Letters B} {\bfseries 773} (2017) 422 }.

\bibitem{vanRees2011}
B.~C. van Rees, \emph{Holographic renormalization for irrelevant operators and
  multi-trace counterterms},
  \href{https://doi.org/10.1007/JHEP08(2011)093}{\emph{Journal of High Energy
  Physics} {\bfseries 2011} (2011) 93}.

\bibitem{Papadimitriou2011}
I.~Papadimitriou, \emph{Holographic renormalization of general dilaton-axion
  gravity}, \href{https://doi.org/10.1007/JHEP08(2011)119}{\emph{Journal of
  High Energy Physics} {\bfseries 2011} (2011) 119}.

\bibitem{Chemissany2015}
W.~Chemissany and I.~Papadimitriou, \emph{Lifshitz holography: the whole
  shebang}, \href{https://doi.org/10.1007/JHEP01(2015)052}{\emph{Journal of
  High Energy Physics} {\bfseries 2015} (2015) 52}.
  
   \bibitem{grozdanov}
  S.~Grozdanov,  and N.~Poovuttikul, \emph{Generalized global symmetries in states with dynamical defects: The case of the transverse sound in field theory and holography} \href{https://link.aps.org/doi/10.1103/PhysRevD.97.106005}{\emph{Phys. Rev. D}{\bfseries 97} (2018) 106005}.

\bibitem{Arav2017}
I.~Arav, Y.~Oz and A.~Raviv-Moshe, \emph{Lifshitz anomalies, Ward identities
  and split dimensional regularization},
  \href{https://doi.org/10.1007/JHEP03(2017)088}{\emph{Journal of High Energy
  Physics} {\bfseries 2017} (2017) 88}.

\bibitem{Hartong2015}
J.~Hartong, E.~Kiritsis and N.~A. Obers, \emph{Field theory on Newton-Cartan
  backgrounds and symmetries of the Lifshitz vacuum},
  \href{https://doi.org/10.1007/JHEP08(2015)006}{\emph{Journal of High Energy
  Physics} {\bfseries 2015} (2015) 6}.

\bibitem{Andrade2014}
T.~Andrade and B.~Withers, \emph{A simple holographic model of momentum
  relaxation}, \href{https://doi.org/10.1007/JHEP05(2014)101}{\emph{Journal of
  High Energy Physics} {\bfseries 2014} (2014) 101}.

\bibitem{Sonner2013}
J.~Sonner, \emph{On universality of charge transport in AdS/CFT},
  \href{https://doi.org/10.1007/JHEP07(2013)145}{\emph{Journal of High Energy
  Physics} {\bfseries 2013} (2013) 145}.

\bibitem{Tong2013}
D.~Tong and K.~Wong, \emph{Fluctuation and dissipation at a quantum critical point},
\href{https://link.aps.org/doi/10.1103/PhysRevLett.110.061602}{\emph{Phys. Rev. Lett.}
{\bfseries 110} (2013) 061602}

\bibitem{Mozaffar2013}
M.~R. M. Mozaffar and A.~Mollabashi, \emph{Quantum critical points in Lifshitz space-time}, 
\href{https://doi.org/10.1007/JHEP04(2013)081}{\emph{Journal
  of High Energy Physics} {\bfseries 2013} (2013) 81}.

\bibitem{Davison2017}
R.~A. Davison, W.~Fu, A.~Georges, Y.~Gu, K.~Jensen and S.~Sachdev,
  \emph{Thermoelectric transport in disordered metals without quasiparticles:
  The Sachdev-Ye-Kitaev models and holography},
  \href{https://doi.org/10.1103/PhysRevB.95.155131}{\emph{Phys. Rev. B}
  {\bfseries 95} (2017) 155131}.

\bibitem{Donos2014}
A.~Donos and J.~P. Gauntlett, \emph{Thermoelectric DC conductivities from black
  hole horizons}, \href{https://doi.org/10.1007/JHEP11(2014)081}{\emph{Journal
  of High Energy Physics} {\bfseries 2014} (2014) 81}.

\bibitem{deBoer2018}
J.~de~Boer, J.~Hartong, N.~A. Obers, W.~Sybesma and S.~Vandoren, \emph{{Perfect
  fluids}}, \href{https://doi.org/10.21468/SciPostPhys.5.1.003}{\emph{SciPost
  Phys.} {\bfseries 5} (2018) 3}

\end{thebibliography}

\end{document}